\documentclass{sn-jnl}

\usepackage{graphicx}%
\usepackage[font=small,labelfont=bf]{caption}
\usepackage{multirow,multicol}%
\usepackage{amsmath,amssymb,amsfonts}%
\usepackage{amsthm}%
\usepackage{mathrsfs}%
\usepackage[title]{appendix}%
\usepackage{xcolor}%
\usepackage{textcomp}%
\usepackage{manyfoot}%
\usepackage{booktabs,pdflscape}%
\usepackage{algorithm,comment}%
\usepackage{algorithmicx}%
\usepackage{algpseudocode}%
\usepackage{listings}%
\usepackage [utf8]{inputenc}

\usepackage{hyperref}

\usepackage[utf8]{inputenc}
\usepackage[T1]{fontenc}

\usepackage{pdflscape,longtable,array,caption}
\begin{document}
\title[]{Geospatial Soil Quality Analysis: A Roadmap for
Integrated Systems}

\author*[1]{\fnm{Habiba} \sur{BEN ABDERRAHMANE\orcid{0009-0008-6212-2106}}}\email{habiba.benabderrahmane@lagh-univ.dz}

\author[2]{\fnm{Slimane} \sur{OULAD-NAOUI}\orcid{0000-0001-6311-5081}}\email{s.ouladnaoui@univ-ghardaia.edu.dz}%
\equalcont{These authors contributed equally to this work.}

\author[1]{\fnm{Benameur} \sur{ZIANI}}\email{bziani@lagh-univ.dz}
\equalcont{These authors contributed equally to this work.}


\affil*[1]{\orgdiv{Laboratoire d'Informatique et de Mathématiques (LIM)}, \orgname{University of Ammar Telidji}, \orgaddress{\street{37G Ghardaia Road}, \postcode{03000}, \state{Laghouat}, \country{Algeria}}}
\affil[2]{\orgdiv{Laboratoire de Mathématiques et Sciences Appliquées (LMSA)}, \orgname{Université de Ghardaia}, \orgaddress{\street{Scientific Zone, PO Box 455}, \postcode{47000}, \state{Ghardaia}, \country{Algeria}}}


\abstract{Soil quality plays a crucial role in sustainable agriculture, environmental conservation, and land-use planning. Traditional SQ assessment techniques rely on costly, labor-intensive sampling and laboratory analysis, limiting their spatial and temporal coverage. Advances in Geographic Information Systems (GIS), remote sensing, and machine learning (ML) enabled efficient soil quality evaluation. This paper presents a comprehensive roadmap distinguishing it from previous reviews by proposing a unified and modular pipeline that integrates multi-source soil data, GIS and remote sensing tools, and machine learning techniques to support transparent and scalable soil quality assessment. It also includes practical applications. Contrary to existing studies that predominantly target isolated soil parameters or specific modeling methodologies, this approach consolidates recent advancements in Geographic Information Systems (GIS), remote sensing technologies, and machine learning algorithms within the entire soil quality assessment pipeline. 
It also addresses existing challenges and limitations while exploring future developments and emerging trends in the field that can deliver the next generation of soil quality systems making them more transparent, adaptive, and aligned with sustainable land management.

}



\maketitle

\section{Introduction}\label{intro}
Soil quality plays a fundamental role in agriculture, environmental sustainability, and land management. It supports food security, ecosystem resilience, and human well-being \cite{sumathi_improved_2023}. The concept of soil quality was initially formalized in the early 1990s, refers to the capacity of soil to perform essential tasks. According to the definition given by  USDA (United States Department of Agriculture) in 1994, soil quality is defined as the capacity of a given soil type, within its inherent or managed ecological context, to preserve or enhance air and water quality, sustain biotic productivity, and support the health of both human populations and ecological systems \cite{Swapana_Sepehya2024}.  

During the past three decades, the concept has evolved from a narrow focus on productivity to a broader recognition of soil’s role in ecological functioning. Soil quality is not directly measurable; instead, it is inferred from a suite of \textit{physical}, \textit{chemical}, and \textit{biological} indicators that respond to environmental change and land management practices. These indicators capture the dynamic nature of the soil and provide the basis for evaluation frameworks.  

Geographic Information Systems (GIS) and remote sensing have become powerful tools to determine the spatial distribution of environmental variables and soil properties \cite{omran_improving_2012, pereira_construction_2022}. In soil science, spatial prediction and surface modeling are now widely used, allowing researchers to examine landscape variability and enhance soil management and precision agriculture decision-making  \cite{sumathi_improved_2023,padarian_using_2019}. However, the accuracy of such maps depends heavily on sampling strategies and interpolation methods, both of which remain active areas of research \cite{omran_improving_2012}.  

Traditionally, soil quality evaluation has relied on manual field sampling and laboratory investigations and analyses that are expensive and take time. As a result, methods utilizing artificial intelligence techniques have increasingly been adopted for predicting soil quality indicators. Various Deep Learning/Machine Learning models have been explored for Soil Quality Prediction each offering unique advantages. 

Despite this progress, much of the existing literature remains fragmented, often focusing on a single property or a limited application resulting in a lack of comprehensive overviews that address GIS-based soil quality systems holistically. To address this gap, this paper proposes a roadmap for developing geospatial soil quality systems as a unified and modular pipeline. The proposed approach integrates diverse data sources, preprocessing workflows, analytical engines, and available tools. Developed through the synthesis of existing methodologies, this pipeline provides a practical blueprint for creating adaptable and transparent GIS-based soil quality management systems. Furthermore, the paper reviews recent applications in precision agriculture, soil contamination, land use change, fertility assessment, land degradation, salinization, and climate change.

The remainder of this paper is structured as follows : in Section \ref{SoilProperties} we presented a foundational
understanding of soil quality properties and indices. Section \ref{gis} is dedicated to the basics of geospatial analysis systems for soil quality analysis, followed by the tools and frameworks available in \ref{AvToolsFmwk}. Section \ref{relatedWorks} reviews related work. Section \ref{challenges} discusses key challenges and limitations. Followed by the trends and future research directions highlighted in Section \ref{trends}. Finally, Section \ref{concl} presents the conclusion.

\section{Soil quality characteristics and indicators}\label{SoilProperties}

To develop effective geospatial soil quality systems, it is important to first establish a clear understanding of the criteria that define soil quality and the methods used for its measurement. The following section presents key soil characteristics and indicators that serve as the foundation for assessment frameworks.


Assessing soil quality requires an integrated understanding of its physical, chemical, and biological properties, each of which reflects the soil's capacity to perform essential ecological and productive functions \cite{DevSQAframework2025}. 

The key \textit{physical indicators} are soil texture, bulk density, porosity, aggregate stability, and infiltration rate. The texture of the soil affects the retention of water, the holding capacity of nutrients and the penetration of the roots. While bulk density is an indicator of compaction and aeration status. These parameters are fundamental in assessing soil structure and hydrological behavior, and can be modeled using Digital Soil Mapping (DSM) techniques combined with terrain attributes derived from Digital Elevation Models (DEMs).

\textit{Chemical indicators} such as pH, electrical conductivity (EC), organic carbon (OC) content, and cation exchange capacity (CEC) provide insight into nutrient cycling, microbial activity, and soil fertility. The availability of macronutrients including nitrogen (N), phosphorus (P), and potassium (K) is especially important for agricultural productivity. 

Equally essential are \textit{biological indicators}, though more difficult to measure at scale, are crucial for understanding the ecological functions of soil. Parameters such as microbial biomass, soil respiration, enzyme activity, and the abundance of soil fauna provide critical information on decomposition processes, nutrient cycling, and soil resilience. 

To facilitate multiparameter analysis, a range of \textit{soil quality indicators} (SQIs) have been developed. These indices synthesize physical, chemical, and biological measurements into a unified metric that reflects soil health status. In geospatial platforms, SQIs are frequently mapped to support applications including precision agriculture, land degradation assessment, and informed policy development \cite{DevSQAframework2025}.  


Having established a comprehensive understanding of soil quality indicators, the subsequent focus is on examining how these indicators are represented, processed, and analyzed through geospatial technologies. Section 3 details the essential components and workflows of geospatial systems used in soil quality analysis.

\section{Geospatial Systems for Soil Quality Analysis: Proposed
Pipeline}\label{gis}
Geospatial data represent where an object is located, i.e., the association with the geometric parameters of the object and the spatial
references or coordinates. 
 Authors of \cite{grekousisSpatialThinkingCambridge2020} define spatial analysis as "a collection of methods, statistics, and techniques that integrate concepts such as location, area, distance, and interaction to analyze, investigate, and explain in a geographic context patterns, actions, or behaviors among spatially referenced observations that arise as a result of a process operating in space".\\

A fundamental concept in spatial analysis is GIS, which is defined as "a computer-based system for the acquisition, storage, management, querying, analysis, and visualization of geospatial data" \cite{changGISintroduction2019}. It was introduced by Dr. Roger Tomlinson, known as the father of GIS, who developed the first GIS system known as the \textit{Canada Geographic Information System} (CGIS), which was used to catalog and analyze Canada's natural resources \cite{lawheadPyHist2019}. 

\begin{figure}[!h]
	\includegraphics[height=4cm,width=\linewidth]{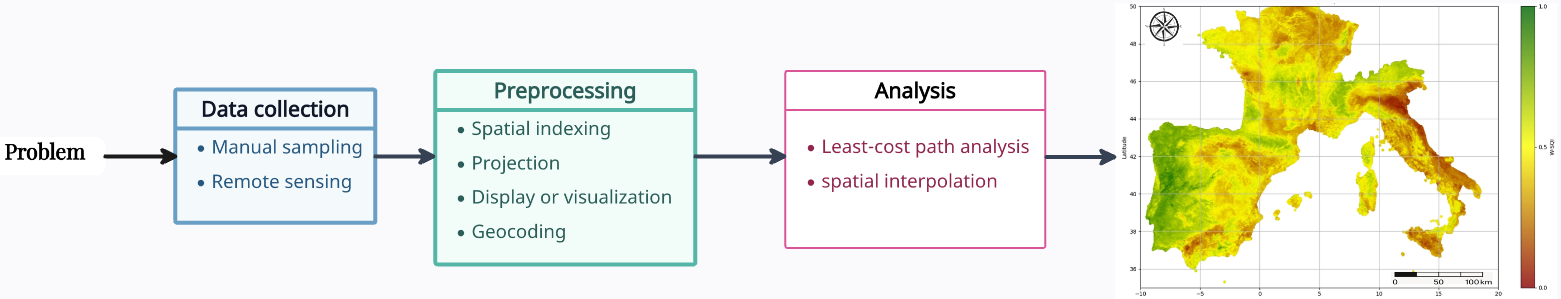}
\caption{Proposed GIS analysis pipeline system.}\label{f:GDApipeline}
\end{figure}

Based on the reviewed literature, this study introduces a unified, modular GIS-based soil quality analysis pipeline, as illustrated in Fig.~\ref{f:GDApipeline}. Based on existing research \cite{changGISintroduction2019, lawheadPyHist2019, lovelaceGeocomputationR2020}, the proposed integrated architecture and workflow constitute an innovative consolidation. The pipeline synthesizes heterogeneous data sources, computational tools, and analytical methodologies into a cohesive system designed for scalable, transparent, and adaptive soil quality evaluation. It functions as a reference architecture to facilitate the development and enhancement of geospatial soil monitoring systems by researchers and practitioners.


\subsection{Data collection and Storage} 

\begin{figure}
	
\centering	\includegraphics[width=7cm]{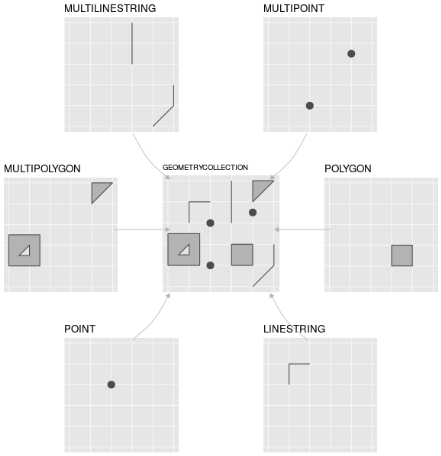}
	
	\caption{Spatial Objects}\label{f:SptiaObjcts}
\end{figure}

\begin{figure}
\centering	\includegraphics[width=10cm]{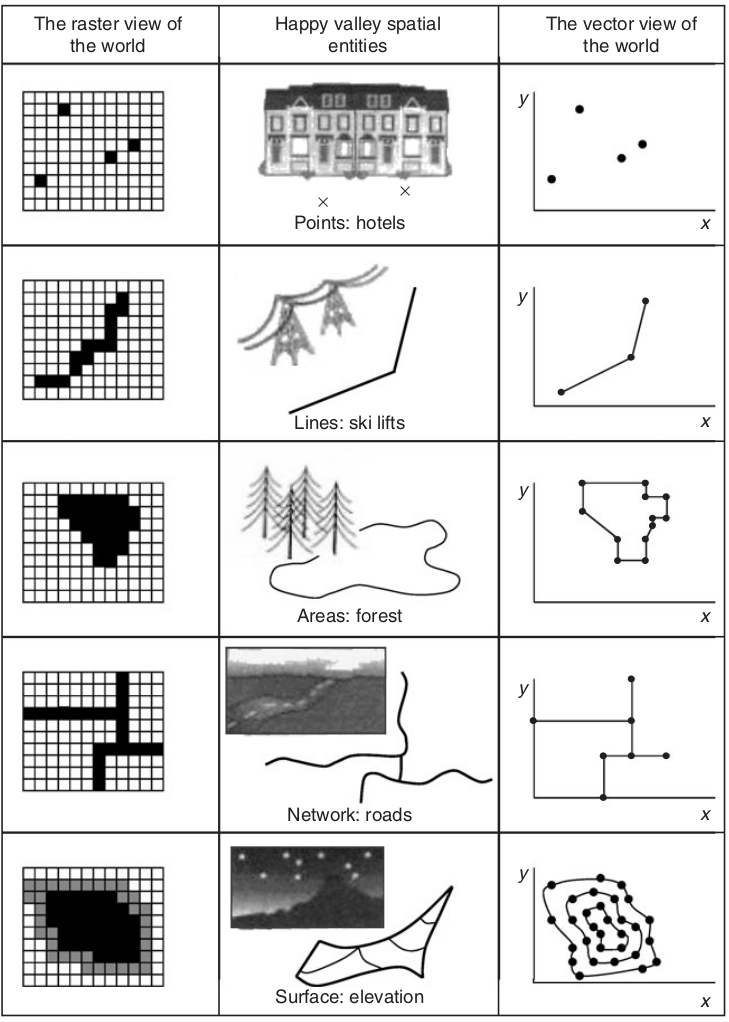}
		\caption{Spatial Objects representation}\label{f:SptiaObjctsVecRast}
\end{figure}

Figure.~\ref{f:SptiaObjcts} illustrates the main types of GIS spatial objects, namely points, lines, and polygons \cite{lovelaceGeocomputationR2020}, that constitute the fundamental geometric components of the geospatial data models employed in soil analysis. 
Precise comprehension of these geometric primitives is critical for the organization of soil datasets and the implementation of suitable spatial analysis procedures within GIS. These objects can be stored as Vectors or Raster, as illustrated by Figure \ref{f:SptiaObjctsVecRast}, where :
\begin{itemize}
	\item \textbf{Vectors} that contain the information about points, lines or polygons, with their specific coordinates.
	\item \textbf{Raster} A raster model subdivides the entire study area into a systematic grid of cells, analogous to pixels in digital remote sensing imagery, with each cell representing a single attribute value~\cite{QihaoRemoteSensing}. 
\end{itemize}
\textcolor{white}{   }

Late studies collected in-situ sampling manually, while recently an available source of geospatial data is the \textit{remote sensing}, which 
refers to the systematic acquisition of information about the Earth's surface, comprising terrestrial and aquatic environments, as well as the atmosphere, via sensors mounted on airborne platforms (e.g., aircraft, balloons) or spaceborne platforms (e.g., satellites, space shuttles)~\cite{QihaoRemoteSensing}. Depending on the application domain and data acquisition methodology, remote sensing modalities are classified into the following categories: (1) satellite remote sensing, which employs orbital sensors; (2) photographic and photogrammetric techniques, which utilize optical imaging through photography; (3) thermal remote sensing, which detects thermal infrared radiation to infer surface temperature; (4) radar remote sensing, such as Synthetic Aperture Radar (SAR), which employs microwave wavelengths for surface imaging; and (5) Light Detection and Ranging (LiDAR), which determines distances by emitting laser pulses and measuring their return time~\cite{QihaoRemoteSensing}.
 
To facilitate the development of comprehensive geospatial soil quality analysis systems, a diverse collection of datasets is available, each differing in scope, resolution, data type, and measured variables, enumerated as follows.
\begin{itemize}
	\item \textbf{AfSIS}\footnote{\url{https://www.isric.org/projects/africa-soil-information-service-afsis}} 
	Soil datasets and maps for Africa. It focuses specifically on sub-saharan Africa, which maps soil properties such as: soil organic carbon, pH, texture (sand, silt, clay), exchangeable nutrients (Ca, Mg, K, etc.), CEC, bulk density.
	
	\item \textbf{European Soil Database (ESDB)}\footnote{\url{https://esdac.jrc.ec.europa.eu/}}  is a comprehensive geodatabase developed by the European Commission's Joint Research Centre (JRC). It provides harmonized soil information for Europe, including a wide range of attributes such as: soil texture, depth to bedrock, organic carbon content, drainage, stoniness, pH, parent material, land use.
	 
	\item  \textbf{LUCAS} (Land Use/Cover Area frame Survey)\footnote{\url{https://esdac.jrc.ec.europa.eu/content/lucas-2009-topsoil-data}} is a large-scale in-situ field survey conducted by Eurostat (the statistical office of the European Union) to collect harmonized data on land use, land cover, and agricultural practices across EU member states. It contains the following properties : soil organic carbon (SOC), pH, texture (sand, silt, clay), bulk density, carbonates, heavy metals.

	\item \textbf{FAO (Food and Agriculture Organization) Global Soil Map}\footnote{\url{https://www.fao.org/soils-portal/data-hub/soil-maps-and-databases/en/}} is a collection of global and regional soil information products developed and maintained by the FAO Organization. It contains \textit{GSOCmaps} (Global Soil Organic Carbon Map), \textit{GloSIS} (Global Soil Information System), \textit{HWSD} (Harmonized World Soil Database), soil pH, texture, salinity, nutrient status properties.
	
	\item \textbf{SoilGrids}\footnote{\url{https://soilgrids.org/}} developed by ISRIC-World Soil Information. It provides detailed, high-resolution maps of soil \textit{properties} such as bulk density, sand, silt, clay fractions, coarse fragments and soil depth, organic carbon, pH, cation exchange capacity; and \textit{classes} based on machine learning and large datasets of soil observations. It generates maps at a spatial resolution of up to 250 meters.  
	
	\item \textbf{Soil Data Mart (USDA NRCS)}\footnote{\url{https://websoilsurvey.nrcs.usda.gov/}} developed by the USDA’s Natural Resources Conservation Service (USDA NRCS) to provide access to official soil survey data across the United States. It has now been replaced by the Web Soil Survey (WSS) and integrated into the Soil Data Access (SDA) system, which continues to provide public access to NRCS-certified soil data and contains soil data: texture, structure, and horizon data, soil organic matter and pH, Hydrologic group and drainage, land capability classes, engineering and building site development interpretations, crop suitability and forest productivity.
	
		
	\item \textbf{MODIS Soil Moisture Data (NASA)}\footnote{\url{https://modis.gsfc.nasa.gov/data/}} based on data from MODIS instruments aboard NASA’s Terra (since 1999) and Aqua (since 2002) satellites. It provides information with spatial resolution (250m to 1km) about soil moisture which is not directly sensed but modeled using indices derived from MODIS, such as: NDVI/EVI (vegetation health), LST (land surface temperature), Albedo, emissivity.
	
	\item \textbf{SIS} \footnote{\url{https://esdac.jrc.ec.europa.eu/}} developed by the Joint Research Centre (JRC) of the European Commission is a comprehensive digital framework that integrates, manages, and disseminates soil data and assessments across Europe. Is Is multi-layered system, including: spatial soil databases (e.g., European Soil Database), thematic soil maps (e.g., erosion, organic carbon, contamination), soil profile datasets, model outputs and indicators. 
	
	\item \textbf{ISRIC - World Soil Information}\footnote{\url{https://www.isric.org/}} is a global soil data platform developed and maintained by International Soil Reference and Information Centre (SRIC). It provides access to a harmonized, curated, and geo-referenced database of soil profile observations from around the world. It offers consistent soil point data for various soil properties : OC, pH (H$_2$O and KCl), texture (sand, silt, clay), bulk density, CEC, base saturation, EC, Nutrients (e.g., P, K, N).
	
	\item \textbf{SWAT Soil Database}\footnote{\url{https://swat.tamu.edu/}} 	 designed specifically for the hydrological model, used worldwide for watershed-scale simulation of water quantity and quality. It contains the following properties : soil depth,t exture (sand, silt, clay), bulk density, available water capacity (AWC), hydraulic conductivity (Ksat), rock fragment content, organic carbon content.
	
	\item \textbf{LandPKS Soil Data}\footnote{\url{https://landpks.org/}} 	 combines field observations with soil science knowledge to assess land potential, soil health, and guide sustainable land management decisions. It empowers users with tools to gather soil profile data directly in the field using smartphones, helping fill data gaps in underserved regions worldwide. The field-measured or observed: Soil texture by feel (sand, silt, clay), Soil color (using Munsell charts), Soil depth and rooting depth, Rock fragment content, Surface and subsurface horizon characteristics.

\end{itemize}

Table~\ref{t:soil_datasets} offers a comparative overview of primary free soil datasets commonly utilized in GIS-based soil quality research and applications.
\begin{table}[!h]
\centering
\caption{Comparative Summary of Major Soil Quality Datasets}\label{t:soil_datasets}
\begin{tabular}{|p{2.cm}|p{2.cm}|p{3.7cm}|p{1.6cm}|p{2.cm}|}
\hline
\textbf{Dataset} & \textbf{Geographic Scope} & \textbf{Key Variables} & \textbf{Data Type} & \textbf{Resolution} \\
\hline
\textbf{SoilGrids (ISRIC)} & Global & Bulk density, sand/silt/clay fractions, organic carbon, pH, cation exchange capacity (CEC), soil depth & Raster (ML-derived) & $\sim$250 m \\
\hline
\textbf{FAO Global Soil Map} & Global and regional & Soil organic carbon (SOC), pH, salinity, texture, nutrient status, HWSD, GloSIS, GSOCmaps & Raster and tabular & Varies by product \\
\hline
\textbf{LUCAS (EU)} & European Union & SOC, pH, texture, bulk density, heavy metals, land use and management data & In-situ point data & Point-based (field survey) \\
\hline
\textbf{European Soil Database (ESDB)} & Europe & Soil texture, depth to bedrock, organic carbon, pH, parent material, drainage, land use & Vector and raster & 1 km \\
\hline
\textbf{MODIS Soil Moisture} & Global (satellite) & Soil moisture (modeled), NDVI, land surface temperature (LST), albedo, emissivity & Raster (remote sensing) & 250 m – 1 km (daily) \\
\hline
\textbf{AfSIS (Africa Soil Information Service)} & Sub-Saharan Africa & pH, SOC, texture, CEC, bulk density, exchangeable nutrients (Ca, Mg, K) & Raster and point data & $\sim$250 m \\
\hline
\textbf{SIS (Soil Information System)} & Europe & Spatial soil databases, erosion, contamination, organic carbon, soil profile datasets and indicators & Multi-layer GIS & Variable \\
\hline
\textbf{USDA Soil Data Mart / Web Soil Survey} & United States & Texture, structure, organic matter, pH, drainage, engineering properties, crop suitability & Vector and tabular & County-scale or finer \\
\hline
\textbf{ISRIC Soil Profiles} & Global & SOC, pH, texture, bulk density, electrical conductivity (EC), CEC, nutrients, base saturation & Geo-referenced point data & Point locations \\
\hline
\textbf{SWAT Soil Database} & Global (hydrological focus) & Texture, bulk density, hydraulic conductivity, organic carbon, available water capacity (AWC), soil depth & Tabular (model input) & Watershed scale \\
\hline
\textbf{LandPKS} & Global (field-level) & Soil texture by feel, color, rooting depth, rock fragment content, surface/subsurface characteristics & Mobile survey and app data & Field scale \\
\hline
\end{tabular}
\label{tab:soil_datasets}
\end{table}
The comparison presented in Table~\ref{t:soil_datasets} delineates the trade-offs among spatial extent, spatial resolution, and data typology across principal soil datasets. Globally distributed, model-generated datasets such as SoilGrids and MODIS offer lower spatial resolution, making them more suitable for macro-scale spatial analyses. Conversely, in-situ datasets like LUCAS and LandPKS deliver higher measurement precision at the field scale, making them well-suited for localized studies and validation of predictive models. Layered or multi-parameter systems such as European
Soil Database (ESDB) and SIS are optimized for integration within GIS frameworks and thematic mapping applications. Specialized datasets, including SWAT, are tailored for hydrological modeling purposes. Selection of an appropriate dataset must consider the spatial analytical scale, specific research requirements, and the intended modeling approach.

The current soil datasets offered significant potential to advance soil quality assessments, while also presented serious challenges such as semantic harmonization, spatial transformation, temporal integration, and bias correction.

Following the acquisition of pertinent geospatial and soil datasets, a sequence of preprocessing procedures must be executed to solve these challenges and ensure readiness for analytical applications. The procedures outlined below are critical to harmonizing heterogeneous data sources and facilitating accurate spatial analysis and modeling processes.

\subsection{Preprocessing}  
Preprocessing is an essential stage in geospatial soil quality analysis systems, acting as a crucial intermediary between data collection and effective analytical modeling stages. Given the variety of data formats, spatial resolutions, and coordinate systems, preprocessing ensures that datasets are accurately cleaned, standardized, and spatially aligned. This process facilitates the integration of multiple sources of geospatial information. The GIS operations include \cite{lawheadPyHist2019} :

\paragraph{Spatial indexing} spatial index represents the way of associating to each data an index to facilitate the access of objects which offers a quick response to queries without examining every entry in the dataset. The most used algorithms are \textit{Quadtree index} and \textit{R-tree index}.

\paragraph{Projection} Since the Earth has a spherical shape, projection is a way to represent the coordinates of an object using a coordinate reference
system (CRS). This representation preserves common properties, such as area, scale, bearing, distance, or shape. Projections are commonly delineated by a collection of more than 40 parameters, presented in either XML or a textual form known as Well-Known Text (WKT), which serves to specify the transformation algorithm. Those WKT are registered by the International Association of Oil and Gas Producers (IOGP) and were formerly known as the European Petroleum Survey Group (EPSG), representing over 5000 entries. For example, the Google Mercator projection used in Google Maps uses the code EPSG:3857.

\begin{figure}
    \centering
    \includegraphics[width=0.5\linewidth]{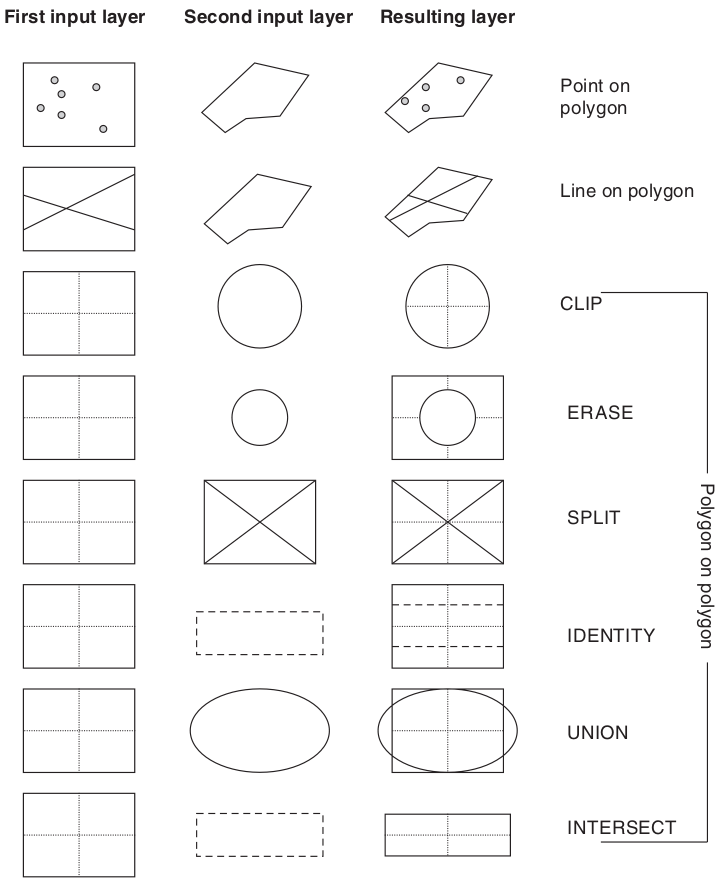}
    \caption{GIS overlay operations}
    \label{f:overlyOperations}
\end{figure}

\paragraph{Overlay operations} Figure \ref{f:overlyOperations} illustrates GIS overlay operations including clipping, intersecting, buffering, merging, dissolving, and erasing, which are used to combine and integrate spatial data layers effectively \cite{korstanjeMLgeoPython2023, QihaoRemoteSensing}.

\paragraph{Geocoding} is the process of converting a street address into latitude and longitude \cite{lawheadPyHist2019}. While \cite{changGISintroduction2019} describes it as an analytical step, we consider it a preprocessing step.

Subsequent to preprocessing, the processed geospatial dataset can be analyzed using a range of spatial analytical methodologies and machine learning algorithms to identify significant spatial patterns, quantify soil properties, and inform decision-making processes. The subsequent section delineates essential spatial analysis techniques and machine learning models employed in soil quality evaluation.

\subsection{Analysis}

The analysis phase is a critical component of geospatial soil quality assessment. This stage generally includes \textit{spatial interpolation}, which is defined as "the estimation of surface values at unsampled points (population) using points with known values (sample)"~\cite{changGISintroduction2019}.

Spatial interpolation methods can be divided in different categories~\cite{changGISintroduction2019}:
\begin{itemize}
	\item \textit{global or local} : Global interpolation refers to methods that utilize the entire dataset of known points to generate estimated values across the domain, exemplified by techniques such as trend surface modeling and regression analysis. In contrast, local interpolation methods employ a subset of known proximal points to interpolate values at target locations. Principal local interpolation techniques encompass  \textit{Thiessen Polygons}, \textit{kernel-based Density Estimation}, \textit{Inverse Distance Weighted (IDW)} interpolation and \textit{Thin-Plate Splines} interpolation.
	\item \textit{exactitude} of interpolation, which is classified based on the property of exactness: exact interpolation methods reproduce the known data values precisely at the data points, whereas inexact (or imprecise) interpolation methods do not guarantee the reproduction of known data values at data points, resulting in estimated values that may differ from the known values.
	\item \textit{deterministic or stochastic}, while stochastic interpolation methods take into account the presence of some unpredictability in their variable and provide evaluation of prediction errors with estimated variances, deterministic interpolation methods do not provide assessment of errors with expected values. Except the Kriging, every previously local techniques are deterministic. 
\end{itemize}

The previous approaches employ mathematical models, where the deterministic methods lack a definitive ground truth~\cite{korstanjeMLgeoPython2023} and the stochastic methods require extensive computational resources~\cite{70YearMLinGeo2020}. Consequently, it is essential to explore alternative techniques that can achieve the same objectives but in a more efficient manner. Thus, the utilization of machine learning techniques becomes relevant.

The aim of ML is to establish computational models that can automatically learn patterns and generalize from input data~\cite{PeterflachML2012}. It offers a collection of algorithms that can handle various types of data (labeled, unlabeled; unstructured, structured, semi-structured; sequential, unsequential) to address crucial challenges: classification
, regression, clustering
, dimensionality reducing
, sequential analysis.

The use of machine learning in the analysis of spatial data was mentioned in the overview of geoscience provided by~\cite{70YearMLinGeo2020} over the last $70$ years. In the $1980$s, Hopfield neural networks were used for seismic deconvolution, which was the first application of artificial neural networks (ANN) in geography. In early 1999, Support Vector Machines (SVMs) were employed for land use classification applications utilizing remote sensing data. While Random Forests demonstrated potential, widespread recognition was limited initially, as the term was not introduced until $2001$ for fault interpretation, facies classification and  seismic interpretation.

For spatial interpolation, an alternative neural network architecture designated as Radial Basis Functions Network (RBFN) is utilized, wherein the activation functions of the hidden layers are Radial Basis Functions (RBFs), originally introduced by Powell (1987) to address multivariate interpolation problems \cite{linChenRBFN2004}. Furthermore, in 2017, a Convolutional Neural Network (CNN) architecture was utilized by A.U. Waldeland and A. Solberg for image-based seismic interpretation \cite{70YearMLinGeo2020}.

The deployment of sequential machine learning architectures, such as Recurrent Neural Networks (RNNs) and Long Short-Term Memory networks (LSTMs), has facilitated substantial progress in spatial data analysis by leveraging diverse existing datasets within the domain. Notable applications, as detailed in \cite{70YearMLinGeo2020}, include a 2017 study employing RNNs to extract geological relationships from unstructured text corpora. Additional applications encompass the classification of seismological events associated with volcanic activity, the prediction of landslide displacement through multivariable modeling, the characterization of sedimentological stratigraphic sequences, and the estimation of petrophysical parameters from seismic attribute data. Furthermore, Generative Adversarial Networks (GANs) have been adopted to synthesize artificial data samples aimed at enhancing automated seismic interpretation \cite{70YearMLinGeo2020}.\\	

Table~\ref{tab:ml_benchmark_soil} provides a comprehensive summary and benchmarking of the machine learning algorithms used in the geospatial evaluation of soil quality, delineating their respective advantages, limitations, and optimal application contexts within this field. 
\begin{table}[htbp]
\centering
\scriptsize
\caption{Benchmark and Comparative Analysis of ML Methods in Geospatial Soil Quality Systems}
\label{tab:ml_benchmark_soil}
\begin{tabular}{|p{2.5cm}|p{3cm}|p{4.5cm}|p{4.5cm}|}
\hline
\textbf{ML method} & \textbf{Optimal application} & \textbf{Advantages} & \textbf{Limitations} \\
\hline
Random Forest (RF) & General soil property prediction, feature selection & - Handles non-linear data well \newline - Robust to noise and outliers \newline - Provides feature importance \newline - Works with tabular + mixed data & - May overfit with small tuning \newline - Slower with large datasets \newline - Less interpretable than linear models \\
\hline
Support Vector Machine (SVM) & Land use classification, soil salinity mapping & - Good for small/medium datasets \newline - Effective in high dimensions \newline - Accurate with clear class boundaries & - Not scalable to big datasets \newline - Kernel selection is critical \newline - Memory and compute intensive \\
\hline
Artificial Neural Network (ANN) & Soil Quality Index prediction, fertility mapping & - Flexible architecture \newline - Captures complex relationships \newline - Works with unstructured data & - Black-box model (low interpretability) \newline - Needs large datasets \newline - Prone to overfitting \\
\hline
Convolutional Neural Network (CNN) & Image-based soil mapping, spectral data analysis & - Excels with spatial and image data \newline - Captures local and global features \newline - Powerful for remote sensing inputs & - Requires large labeled datasets \newline - GPU intensive \newline - Complex tuning \\
\hline
Extreme Gradient Boosting (XGBoost) & High-accuracy fertility prediction, tabular data regression & - Fast and efficient \newline - Handles missing data \newline - Built-in regularization \newline - Often best-in-class performance & - Harder to interpret than RF \newline - Hyperparameter tuning needed \newline - Somewhat a black box \\
\hline
RNN / LSTM & Time-series modeling (e.g. salinity trends, degradation) & - Models sequential soil/climate data \newline - Captures temporal dependencies \newline - Useful for long-term prediction & - Difficult to train (vanishing gradients) \newline - High compute cost \newline - Underused in geospatial tasks \\
\hline
\end{tabular}

\end{table}

\subsection{Evaluation metrics}
To evaluate the performance and robustness of techniques employed in geospatial soil quality analysis, appropriate quantitative metrics must be implemented. These metrics are critical for model comparison, algorithm selection, and the assessment of spatial prediction accuracy across datasets. Metric selection should align with the specific machine learning task (e.g., classification, or spatial interpolation) and consider the spatial characteristics inherent in the data. The metrics are as follows~\cite{sumathi_improved_2023}:

\paragraph{Regression or spatial interpolation Metrics}
Having $ y_i\text{, }\hat{y} $ representing actual and predicted values respectively, the metrics are : 
\begin{itemize}
	\item Root Mean Square Error (RMSE) : Measures the average magnitude of error between actual and predicted values. It is computed by the formula \ref{rmse}:
	\begin{equation}
		\mathrm{RMSE} = \sqrt{ \frac{1}{n} \sum_{i=1}^{n} (y_i - \hat{y}_i)^2 }\label{rmse}
	\end{equation}
	It is used when large prediction errors are unacceptable. 
	\item R$^2$ (Coefficient of Determination) : Indicates how well the model explains the variability in the observed data. A closer to $1$ values of R$^2$ mean a better fit. It is given by the formula~\ref{r2}. \begin{equation}
		R^2 = 1 - \frac{ \sum_{i=1}^{n} (y_i - \hat{y}_i)^2 }{ \sum_{i=1}^{n} (y_i - \bar{y})^2 }\label{r2}
	\end{equation}
	It is used when all errors are equally important; good for comparing model simplicity.
	\item Mean Absolute Error (MAE) : Calculated the average absolute difference between predicted and actual values. It is given by the formula~\ref{mae}. \begin{equation}
		\mathrm{MAE} = \frac{1}{n} \sum_{i=1}^{n} \left| y_i - \hat{y}_i \right| \label{mae}
	\end{equation}
	It is used to measure goodness of fit for regression models.
\end{itemize}

\paragraph{Classification Metrics}
Used when the model predicts categories or classes :
\begin{itemize}
	\item \textit{Accuracy} It's the most basic performance metric, which represents the proportion of correct predictions over total predictions. It is computed using the formula~\ref{acc}. \begin{equation}
		\mathrm{Accuracy} = \frac{TP + TN}{TP + TN + FP + FN}\label{acc}\end{equation}
	\item \textit{Precision} represents the proportion of true positive predictions out of all positive predictions made. It is used when false positives are costly, its formula is given by~\ref{precision}.
	\begin{equation}
		\mathrm{Precision} = \frac{TP}{TP + FP}\label{precision}
	\end{equation}
	\item \textit{Recall (Sensitivity)} is the proportion of actual positives that were correctly identified. It is used when missing a true case is dangerous. 
	\begin{equation}
		\mathrm{Recall} = \frac{TP}{TP + FN}
	\end{equation}	
	\item \textit{F1-Score} is a harmonic mean of precision and recall. It balances false positives and false negatives, used when both precision and recall are important in classification. It is given by the formula~\ref{f1s}.
	\begin{equation}
		F_1 = 2 \cdot \frac{\mathrm{Precision} \cdot \mathrm{Recall}}{\mathrm{Precision} + \mathrm{Recall}}\label{f1s}
	\end{equation}

\end{itemize}
Where TP is True Positive, the number of items that are  classified in its proper class; TN is True Negative represents cases where the model accurately predicts the absence of a class; FP is False Positive occurs when the model incorrectly classifies a negative case as positive, leading to a Type I error. and FN extends for False Negative appears when the model fails to detect a condition that is actually present, constituting a Type II error.

After getting an overview about the components of a GIS-based system,  the below section is dedicated to the available tools and frameworks that allows to create that systems.


\section{Available tools and frameworks}\label{AvToolsFmwk}

To create a GIS-based soil quality analysis system, numerous tools are available. This section provides an overview of frequently used tools.

\paragraph{GIS}

GIS Software has become essential for supporting spatial analysis and informed, data-based decision-making. Options ranging from open-source solutions like QGIS\footnote{\url{https://www.qgis.org/}} and GRASS GIS\footnote{\url{https://grass.osgeo.org/}} to commercial platforms such as ArcGIS\footnote{\url{https://www.esri.com/en-us/arcgis/geospatial-platform/overview}} and MapInfo\footnote{\url{https://www.precisely.com/product/precisely-mapinfo/mapinfo-pro/}} Professional offer comprehensive features suited to wide user applications.

\paragraph{Remote Sensing Technologies} As given by \cite{QihaoRemoteSensing} it can be categorized as:
\begin{itemize}
	\item \textbf{Landsat satellites} (Landsat 8, Landsat 9) provide high-resolution multispectral imagery (30m spatial resolution) covering a wide range of the electromagnetic spectrum (visible, near-infrared, shortwave infrared).
	\item \textbf{Sentinel satellites}, part of the \textit{Copernicus} program, provide high-resolution imagery (10m spatial resolution) across 13 spectral bands, from visible to shortwave infrared.
\end{itemize}

\paragraph{Big Data integration}
The creation of general and multipurpose systems requires a complex and large scale data from different sources and computational resources. Geospatial data analysis can be performed using Big data tools, such as Hadoop-based frameworks, such as SpatialHadoop\footnote{\url{https://spatialhadoop.cs.umn.edu/}} and HadoopGIS\footnote{\url{https://github.com/bunnyg/Hadoop-GIS}}, or Spark-based frameworks, including GeoTrellis\footnote{\url{http://geotrellis.github.io/}} and GeoSpark\footnote{\url{https://github.com/agile-lab-dev/GeoSpark}}. Additionally, other proposed solutions, such as "\textit{BiGeo}"\cite{Liu2019}, may also be utilized depending on the specific application requirements.

\paragraph{Cloud solutions}
While big data frameworks require extensive computational infrastructure, cloud platforms provide scalable storage and processing capabilities, enabling efficient execution of large-scale data processing via distributed computing architectures. These platforms support remote access, enhance collaborative efforts among researchers and institutions across diverse geographic locations, and facilitate resource sharing. Several cloud-based solutions are available for geospatial data storage and analysis, including \textit{Alibaba Cloud}, \textit{Amazon Web Services} (AWS), \textit{GeoCloud}, and \textit{Google Earth Engine} \cite{Liu2019}.

\noindent Having analyzed the underlying technological architectures of geospatial information systems, we now investigate their practical applications in domains such as Agricultural Productivity, Environmental Risk and Pollution, Land Use and Degradation, and Climate Interactions.

\section{Applications}\label{relatedWorks}
	
    
    Applications of geospatial soil quality analysis encompass a wide range of sectors, including agriculture and environmental risk management. Using GIS, remote sensing, and machine learning technologies, these approaches facilitate real-time decision-making and predictive modeling across various spatial scales. 

    Recent research has emphasized the importance of integrating soil science knowledge into machine learning frameworks to improve the reliability and interpretability of soil predictions. \cite{Minasny2024} introduced the concept of Soil Science-Informed Machine Learning (SoilML), which embeds domain knowledge into data-driven models through observational priors, soil-specific model architectures, and science-guided loss functions. This framework strengthens predictions in applications such as soil fertility mapping, spectroscopy, and pedotransfer functions, while also addressing issues of interpretability and uncertainty that limit conventional "black-box" approaches.
	
    To illustrate the practical applications of geospatial soil quality systems, this section reviews recent implementations across various sectors. The subsequent subsections are organized by thematic area for clarity and focus. These categories include:
        \begin{itemize}
            \item Agricultural Productivity : focuses on Precision Agriculture and Soil Fertility Analysis
            \item Environmental Risk \& Pollution : addresses Soil Contamination Analysis and Salinization threats
            \item Land Use \& Degradation : investigates land use changes and degradation patterns.
            \item Climate Interaction \& Adaptation : explores how soil data support climate modeling and mitigation.
        \end{itemize}
        The current sections provide detailed discussions with relevant studies and technological methodologies.
    
	\subsection{Precision Agriculture} Precision agriculture utilizes geospatial data analysis to optimize crop yields and reduce soil degradation. By analyzing assembling information's layers of chemical and biological properties related to soil fertility, moisture content, and topography, to produce a map showing which factors influence crop yield. Using remote sensing big data allows the evaluation of amount of fertilizers or pesticides to use, the estimation of productivity and crop yields, the assessment of appropriate crop types or irrigation needs~\cite{Soil}.
	
	Ramzan et al. \cite{Ramzan2024} proposed a two-module crop recommendation system, integrating environmental and soil parameters to guide crop selection. The first module utilizes a combination of static data, such as publicly available datasets and machine learning algorithms including K-Nearest Neighbors (KNN), Decision Tree, RF, SVM, LightGBM, CatBoost, and Adaptive Boosting (AdaBoost). This module was trained on the Kaggle ``Crop Recommendation Dataset'', which contains approximately 100 samples for each of 22 crop types. The second module incorporates real-time data from Internet of Things (IoT) sensors deployed in agricultural areas of Punjab, Pakistan, focusing on six crops (wheat, rice, cotton, maize, gram, and groundnut), supplemented by manual inputs from local farmers.
	
	While precision agriculture aims to optimize input efficiency and maximize crop yields, these objectives rely on a comprehensive understanding of soil fertility. The subsequent section will explore how geospatial systems evaluate and map soil nutrient and fertility levels to support sustainable land management practices.

	\subsection{Soil Fertility Analysis}
	Soil fertility analysis is an essential practice for sustainable agricultural management, enabling farmers and land planners to make informed decisions regarding soil health and nutrient management. Various studies have employed GIS and other geo-spatial techniques to create detailed soil fertility maps, which assess the spatial distribution of soil nutrients and properties. These maps serve as critical tools for optimizing crop production and ensuring environmental sustainability.
	
	
	Several recent studies have developed models to assess and predict soil quality and fertility using geospatial and machine learning approaches. For instance, \cite{Abdellatif2021} proposed the Developed Soil Quality Model (DSQM) to evaluate soil quality in Wadi Al-Halaazin, Matrouh Governorate, Egypt, by integrating GIS, remote sensing data (Sentinel-2, NDVI), digital elevation models (DEM), and field sampling of 48 soil profiles, with ordinary kriging used to interpolate soil parameters such as Fertility Index, Physical Index, Chemical Index, and Geomorphologic Index. In \cite{peng_new_2022}, soil fertility prediction was achieved by selecting optimal crop spectral variables using an extreme gradient boosting (XGBoost) algorithm combined with the variance inflation factor (VIF), followed by fertility mapping through a backpropagation neural network.
	
	Furthermore, \cite{folorunso_exploring_2023} examined supervised machine learning algorithms, specifically, Random Forest (RF), Support Vector Machines (SVM), and Deep Neural Networks (DNN) for quantitative prediction of soil nutrient properties within precision agriculture. The research emphasized advancements in DSM methodologies and demonstrated the integration of Diffuse Reflectance Infrared Spectroscopy (DRIS) with Convolutional Neural Networks (CNN) for spectral data analysis, applying this combined approach to the LUCAS soil dataset to quantitatively estimate six soil chemical parameters.
	
	On the other hand, \cite{el_behairy_accurate_2024} developed a Soil Quality Index (SQI) model that uses 16 physical, chemical, and fertility soil attributes derived from laboratory analyzes, employing an Artificial Neural Network (ANN) implemented in MATLAB, without incorporating geospatial data. 
    
    In our previous study~\cite{mine_comparative_2025}, we compared four machine learning algorithms (RBFN,LightGBM,DNN,XGBoost) to identify the most effective regression to predict soil quality index using the SoilGrid dataset in four European countries (Portugal, Spain, France and Italy. Based on best performance, the XGBoost was selected to generate \textit{soil quality maps}.	


    In addition to nutrient availability, soil health is frequently compromised by exogenous pollutants and toxic substances. The subsequent section examines the application of geospatial analysis and machine learning methodologies for the detection, quantification, and remediation of soil contamination.
    
	\subsection{Soil Contamination Analysis}
	Soil contamination presents a significant threat to environmental integrity and public health, requiring robust assessment and remediation strategies. Analytical techniques offer rapid, precise quantification and spatial characterization of contaminant concentrations, including potentially toxic elements (PTEs), petroleum hydrocarbons (PHCs), and microplastics.
	
	Soil contamination assessment is critical for understanding environmental health risks and managing land resources effectively. Recent advancements in remote sensing and geospatial techniques have significantly enhanced the ability to evaluate soil contamination across various contexts, including industrial areas and mining regions.

	In \cite{Turky_soil_2021}, in situ sampling was performed on 79 soil specimens from the Alpu Plain in Central Anatolia, Turkey, to quantify soil contamination levels utilizing the Enrichment Factor (EF) and Geoaccumulation Index (Igeo). The analytical approach incorporated Principal component analysis (PCA) and Pearson correlation coefficient analysis to elucidate the interdependencies among the identified variables.
	
	Similarly, \cite{Egypt_contamnation_2022} employed ordinary Kriging (OK) for the generation of spatial distribution maps for six heavy metals (Cr, Co, Cu, Cd, Pb, Zn). PCA and contamination factor (CF) calculations were utilized to assess soil contamination severity in the El-Moheet drainage basin, located on the western bank of the Nile River in El-Minia Governorate, Egypt.
	
	An ongoing study of Lat et al.\cite{Malysia_Contamination_soil_2024} 
    performed a detailed geochemical assessment of soil contamination within the Pasir Gudang industrial region of Johor, Malaysia, aiming to characterize the spatial distribution and contamination intensity of selected heavy metals across diverse land-use categories. Topsoil samples (0–20 cm depth) were systematically collected from twenty sites representing industrial, residential, riparian, and educational zones. The samples were subjected to quantitative analysis for major heavy metals, specifically copper (Cu), chromium (Cr), nickel (Ni), zinc (Zn), and lead (Pb), utilizing inductively coupled plasma mass spectrometry (ICP-MS). Contamination levels were quantitatively evaluated using the geoaccumulation index ($I_{geo}$) and contamination factor ($C_f$) metrics. The results indicated that most heavy metal concentrations were below established guideline limits, suggesting the area is generally uncontaminated or only slightly contaminated. The values of $I_{geo}$ and $C_f$ supported these findings, indicating low levels of contamination and minimal anthropogenic impact on soil quality.  However, marginally higher concentrations were detected near certain industrial operations, reflecting localized pollution. The study concluded that current soil management practices in Pasir Gudang were effective, but emphasized the need for ongoing monitoring is essential to detect and mitigate potential pollutant accumulation in industrialized sectors. It is imperative to note that this research methodology does not encompass the application or integration of sophisticated statistical modeling or ML algorithms in its analysis.
	
	A recent study by \cite{Portugal_contaminationAgroSoil_2025} introduced an optimized methodology for the spatial assessment of heavy metal contamination risk, circumventing the logistical constraints associated with physical soil sampling. The study formulated a Probabilistic Pollution Index (PPI) through the integration of GIS technologies with an eight-parameter probability-risk matrix. The parameters incorporated into the model included proximity to roads, industrial sites, soil pH, soil organic matter content, terrain slope, soil texture, proximity to mining areas, and drainage potential. Each parameter underwent a classification and reclassification process to generate a contamination risk surface, whereby individual pixels were categorized into five distinct risk tiers. The efficacy of the PPI was evaluated through comprehensive data analysis involving parameter classification, reclassification procedures, and risk mapping validation.

    Soil contamination primarily results from the ingress of hazardous chemicals and pollutants; however, an equally significant threat to soil integrity stems from natural and anthropogenic processes such as soil salinization, which adversely affects land productivity and disrupts ecosystem equilibrium. The ensuing section delineates geospatial techniques employed for the detection, quantification, and remediation of soil salinity within diverse terrestrial environments.

   	\subsection{Soil Salinization}
	Soil Salinization is a critical global environmental challenge, results from the accumulation of soluble salts in the soil profile, adversely impacting soil health, crop productivity, and ecosystem sustainability. It stems from natural causes such as climate change, sea level rise; and human actions such as poor irrigation practices, excessive water use~\cite{fao2024}. 
	
	Recent studies have extensively investigated soil salinization and degradation using remote sensing, machine learning, and integrative land management approaches. In \cite{he_spatiotemporal_2023}, changes in soil salinization over two decades (2001–2021) in the Werigan–Kuqa River Delta Oasis, Xinjiang, China, were assessed using machine learning models such as Random Forest, LightGBM, Gradient Boosting Decision Trees (GBDT), and XGBoost. Projections to 2050 indicated that, without intervention, salinization could affect up to 50\% of the cultivated land, underscoring the need for targeted management strategies in saline-affected areas.
	
	In response to increasing salinity threats, \cite{tarolli_soil_2024} proposed a hybrid strategy combining nature-based solutions (NBS) such as reforestation, vegetative buffers, and improved land and water management, with bioengineering techniques, specifically the development of salt-tolerant crop varieties through both traditional breeding and genetic engineering. This integrative approach aims to strengthen agricultural resilience, reduce soil degradation, and support adaptation to saline environments.
	
	A long-term retrospective study in Portugal \cite{ramos_soil_2024} examined the causes and evolution of soil salinization over three decades, with a focus on semi-arid and irrigated regions. The study emphasized the influence of climate, soil properties, and land use, and highlighted recent advancements in measurement and modeling techniques. These included the use of pedotransfer functions, remote and proximal sensing technologies, and hydrological modeling tools to better understand salinity dynamics in the vadose zone. The authors advocated for improved fine-scale data, greater integration of soil heterogeneity, and more research on climate change impacts on salinization processes.
	
	Similarly, \cite{li_spatiotemporal_2024} devised a methodology for salinization monitoring in arid regions utilizing multispectral Landsat imagery spanning $2001$ to $2021$ in the Yutian Oasis, China. The study quantitatively assessed the classification accuracy of SVM and Classification and Regression Tree (CART) algorithms for high-resolution spatiotemporal delineation of soil salinity distribution, providing robust analytical tools to inform salinization mitigation strategies and promote sustainable agricultural land use planning.
	
	Furthermore, , in the study conducted by \cite{hagage_monitoring_2024} investigated soil salinization and hydric processes in the northeastern Nile Delta of Egypt, aiming to identify key controlling factors and evaluate soil salinity intensity. The researchers employed hydrochemical analyses of groundwater, irrigation water, and soil samples; utilized water quality indices such as the Irrigation Water Quality Index (IWQI); performed statistical correlation analyses; and integrated remote sensing data from Sentinel-2 imagery and SRTM elevation models to generate spatial maps of waterlogging and salinity patterns using the Inverse Distance Weighting (IDW) interpolation method. The study concludes that the combined effects of poor irrigation water quality, shallow saline groundwater, and restrictive soil textures significantly contributed to severe soil salinity and degradation. These findings impact agricultural productivity and archaeological preservation. The authors recommend urgent mitigation by improved drainage, water table control, and better irrigation practices.
	
	Soil salinization is often correlated with changes in land use and land cover (LULC). The following subsection dedicated for detecting and quantifying LULC transitions.

	\subsection{Land Use Land Cover Change}
	LULC change constitutes a critical domain within spatial analysis, emphasizing the quantification and characterization of landscape modifications resulting from anthropogenic activities and natural dynamics. This process is influenced by a range of drivers, including urban expansion, agricultural intensification, and socio-economic development. 

    Numerous research efforts have employed remote sensing imagery and geospatial analytical methods to systematically detect, monitor, and evaluate LULC transitions across various geographic regions. In \cite{shakya_land_2023}, land use and cover changes in the Guntur district of Singapore were analyzed from 2014 to 2021 using PCA, K-means clustering, and the Multivariate Alteration Detection (MAD) algorithm. Their findings highlighted the impacts of water scarcity on cropping patterns, land use dynamics, rainfall variability, biodiversity risks, and industrial development, with notable changes observed in horticultural land, built-up areas, harvested land, and wasteland.
	
	Similarly, \cite{gubila_land_2024} conducted a spatiotemporal analysis of land use/land cover (LULC) change dynamics in the Semen Bench District of southwest Ethiopia over the period extending from 1986 to 2018. They assessed the consequential alterations in soil physico-chemical attributes associated with these land cover changes. The methodology used involved the acquisition of multispectral Landsat satellite imagery, which was subsequently subjected to geospatial data preprocessing and classification within ERDAS IMAGINE software. A supervised classification approach was executed utilizing the maximum likelihood classifier (MLC) algorithm for each temporal epoch (1986, 2001, 2018), facilitating the generation of spatially explicit LULC classification maps across these defined temporal snapshots. 

	In \cite{brito_land_2025}, researchers analyzed LULC changes in the Portal do Sert\~{a}o situated in Bahia State, Brazil, spanning the temporal interval from 1985 to 2022. The primary objective was to quantify and characterize the nature of spatial transformations and their underlying factors. The methodological framework integrated advanced remote sensing (RS) data acquisition using MapBiomas Collection 8.0 dataset, and GIS to reclassify the land‐cover into four categories: Forest Formation, Agriculture, Urban Area and Water Bodies. These classifications were delineated across discrete temporal snapshots corresponding to the years  1985, 1990, 1995, 2000, 2005, 2010, 2015, 2020, 2022. Their results revealed significant reductions in agricultural and livestock-associated land extents, concomitant with  dramatic increase of urbanized zones and expansion of water surfaces. The spatiotemporal variability of forest cover manifested as episodic fluctuations and were negatively impacted by urbanization and agricultural pressure. The authors inferred that the integration of remote sensing datasets with GIS analytical tools proves efficacious in tracking and modeling spatial-temporal LULC modifications, thereby furnishing critical spatial datasets to inform territorial planning initiatives and environmental policymaking within the region. The investigation utilized categorical spatial analytical methodologies, specifically employing transition matrix formulations and quantification of areal modifications, supplemented by descriptive statistical summaries. No implementation of inferential statistical hypothesis testing, spatial interpolation techniques, or machine learning algorithms was conducted within the scope of the study.
	
	Furthermore, \cite{kucuk_analysis_2025} conducted a comprehensive spatio-temporal analysis of land cover and land use changes across seven geographic regions of the Republic of T\"urkiye, using the CORINE (Coordination of Information on the Environment) land cover classification datasets corresponding to the temporal points of 1990, 2000, 2006, 2012, and 2018. The primary objective was regional differences in land transformation patterns quantify the extent of change within principal land cover classes including artificial surface, agricultural land uses, forested and semi-natural vegetation, wetlands, and surface water bodies. The pipeline processed and analyzed the CORINE data using GIS to generate regional land-cover maps and compute the areal extent and percentage change of each category over time using quantitative metrics, such as area extent (in km²) and percentage change relative to baseline years. The analytical outcomes indicated statistically significant expansions in artificial and urbanized land cover categories, 
    concomitant with a discernible attrition of forested and semi-natural ecosystems across the majority of the study area. Agricultural lands exhibited variable trends, with notable conversions to built-up zones and industrial areas. 
    The study relied on GIS-based spatial quantification and descriptive analysis and did not employ statistical inference, interpolation, or machine learning methods.

	In the study conducted by \cite{senadi_spatio-temporal_2025}, the researchers aimed to delineate and quantify the spatiotemporal changes from agricultural to urban landscapes in the Mitidja Plain of Algeria. They employed multitemporal spectral data derived from Landsat and Sentinel-2 satellite datasets, covering the temporal interval from the year 1983 through 2023, which were processed and analyzed via the Google Earth Engine geospatial analysis platform. Two classification algorithms are investigated, including RF and Smile CART (Classification and Regression Trees). An evaluative comparative analysis indicated that the RF classifier demonstrated superior performance metrics, including enhanced classification accuracy, in detecting and delineating land cover transition zones marked by urban expansion subsequent to agricultural land use. 
    The quantitative results of their analyses revealed a persistent reduction in the extent of cultivated agricultural lands corresponding to an approximate total loss of 28.04\% of initial cultivated land area. Conversely, urbanized, built-up land cover exhibited an increasing by 25.67\% relative to initial urban extents during the same time frame. 
    
	Land use alterations function as principal factors influencing land degradation dynamics. The subsequent section delineates the severity of land degradation and identifies high-risk zones through comprehensive integrated geospatial analysis.
	
	\subsection{Land Degradation Assessment}
	Land degradation~\cite{Barman2020} refers to the decline in land quality and productivity, primarily due to \textit{natural processes}, like erosion, salinization, and drought; \textit{human activities} such as deforestation, overgrazing, and poor irrigation practices. It affects soil health, vegetation cover, water resources, and overall land functionality. It is categorized in four types : \textit{physical degradation}(soil compaction, erosion, crusting), \textit{chemical degradation} (salinity, sodicity, nutrient depletion), \textit{biological degradation} (loss of vegetation cover, biodiversity loss) and \textit{anthropogenic degradation} (unsustainable agriculture, urban sprawl, deforestation).
	
	Numerous studies have investigated land degradation (LD) using remote sensing and geospatial techniques across different regions. For example, \cite{Oroud2023} assessed LD in Jordan's rainfed areas from 1998 to 2021 by analyzing Landsat-derived physical and biophysical metrics, such as the Normalized Difference Vegetation Index (NDVI) and surface albedo. The study found that LD intensifies with increasing aridity, particularly in marginal climate zones, and identified surface albedo as a more stable and reliable indicator than NDVI for monitoring degradation in semiarid environments. Additionally, LD was shown to exacerbate surface temperatures, increase evapotranspiration, and raise forest fire risks, underscoring the value of satellite metrics in developing national LD maps, especially in data-scarce regions.

	An alternate investigation, \cite{Blanche2024} evaluated land cover degradation caused by semi-mechanised and artisanal mining activities in the Mbale, region of Cameroon. This assessment used an integrated remote sensing and digital photogrammetric approach, utilizing multispectral imagery acquired from Sentinel-2 satellite sensors for the years 2019, 2021, and 2023. The collected satellite data underwent pre-processing workflows that included atmospheric correction and seasonal normalization through the computation of spectral vegetation and moisture indices including the NDVI, Normalized Difference Water Index (NDWI), Brightness Index, and Soil Crust Index, to mitigate confounding effects of atmospheric variability and phenological cycles. Subsequently, supervised classification algorithms, namely, Maximum Likelihood Classification (MLC) were implemented to categorize land cover classes, which encompassed vegetative cover, bare soil substrata, hydrological features, and anthropogenic disturbed zones indicative of mining activity. This research underscores the efficacy of combining remote sensing-derived spectral indices, advanced photogrammetric techniques, and field validation protocols to monitor the environmental repercussions of mining activities. Moreover, it contributes to the empirical foundation necessary for informing sustainable land management strategies in environmentally sensitive, mining-impacted landscapes.

	Similarly \cite{Ali2025} conducted a comprehensive land degradation assessment in Egypt's Damietta Governorate, located in the Nile Delta with a Mediterranean climate. Using a multi-index geospatial approach that combined physical, chemical, topographic, wind erosion, and vegetation quality indices, alongside remote sensing, GIS, and ordinary kriging interpolation, the study mapped soil degradation risks. Results indicated that while a large portion of the land remains moderately suitable for agriculture, approximately 32\% faces high to very high degradation risk, mainly due to wind erosion and sparse vegetation. The study emphasizes the need for targeted soil conservation and vegetation restoration efforts, demonstrating the effectiveness of integrating multiple indices and geostatistical techniques for reliable LD mapping. However, the research is limited by its cross-sectional design and the lack of socio-economic factors in its analysis.

    Land degradation is interconnected with climate systems and is significantly modulated by climate dynamics. The following subsection examines the role of geospatial soil system data in climate change modeling, risk assessment methodologies, and adaptation strategy development.

	\subsection{Climate Change} Predictive modeling of climate change can be achieved by analyzing soil properties in conjunction with meteorological data and pollution levels to facilitate early warning systems and inform governmental response strategies~\cite{SoilChallenges, ChineseClimatePrediction}.

    Extensive research efforts have been directed toward understanding climate change phenomena. For instance, \cite{lisetskii_geoinformation_2024} employed geoinformation systems (GIS) and remote sensing technologies to analyze climatic variations across three temporal segments: two $30-$year climatic normals, 1961-1990 and 1991-2020, for the Crimean Peninsula, Ukraine. Their methodology involved examining correlations between climatic variables, such as temperature and precipitation, and soil formation processes to evaluate the influence of climate on soil typology and spatial distribution within the region. By integrating multi-temporal climatic and environmental datasets, the researchers developed high-resolution spatial models elucidating soil property variability as a function of climatic dynamics. The results yield critical insights and provide a foundation for assessing soil renaturation efficacy and the implementation of rehabilitative agricultural practices. 
	Similarly, \cite{yilmaz_trend_2024} investigated the long-term trends in climatic parameters in Sivas Province, Turkey, over a period of 40 years ($1982-2021$). Using GIS, they analyzed climate data to identify shifts in key climatic variables such as temperature and precipitation. The findings provide valuable insights into the climate dynamics of the region, with implications for agriculture, water resources, and land use planning. 		
	\cite{piyasena_utilizing_2024} focuses on using geospatial tools to assess the vulnerability of the Ratnapura District in Sri Lanka to the impacts of climate change. They combined GIS and remote sensing technologies to analyze environmental factors such as temperature variations, precipitation patterns, and land use changes that influence the district's susceptibility to climate-related risks, such as flooding, landslides, and droughts. The findings offer valuable insights for local authorities and planners, enabling them to prioritize areas for climate adaptation and disaster risk reduction measures.

	Several monitoring systems are proposed, such as \cite{gabriele_combined_2023} to monitor land degradation related to climate change in the Basilicata region of southern Italy. Using GIS-based MEDALUS modeling and remote sensing indices (NDVI, albedo) to map climate-related land degradation sensitivity in Basilicata, Italy. They applied multi-criteria index integration and descriptive statistics to provide, for policymakers and local authorities, valuable data and visualized tools for landscape preservation and sustainable land management, planning and environmental conservation.
	
	The GEORES project \cite{lafortezza_geores_2024}, funded by the Italian Space Agency (ASI), is a monitoring system focused on developing a geospatial application aimed at improving environmental sustainability and climate change resilience in urban areas. The project utilizes advanced Earth Observation (EO) technologies, Artificial Intelligence (AI), and eXplainable AI (XAI) to analyze and address risks related to land degradation. It is organized into four primary modules: \textit{Sediment Connectivity}, \textit{Land Displacement}, \textit{Urban Floods}, and \textit{Urban Wildfires}. 

Table \ref{tab:application_comparison} provides a comparative analysis of selected scholarly articles that explore the application of geospatial techniques in soil quality analysis, with a focus on Application Domain Specification, Objective, Methodology / Tools, Scale and Characteristic.

\begin{landscape}
\scriptsize
\begin{longtable}{|p{2.5cm}|p{2.8cm}|p{3cm}|p{3.5cm}|p{1.5cm}|p{3.5cm}|}
\caption{Recapitulation of application papers in geospatial soil quality analysis systems}
\label{tab:application_comparison} \\

\hline
\textbf{Application Domain Specification} & \textbf{Paper} & \textbf{Objective} & \textbf{Methodology / Tools} & \textbf{Scale} & \textbf{Characteristic} \\
\hline
\endfirsthead

\hline
\textbf{Application Domain Specification} & \textbf{Paper} & \textbf{Objective} & \textbf{Methodology / Tools} & \textbf{Scale} & \textbf{Characteristic} \\
\hline
\endhead

\hline
\endfoot

Precision Agriculture & Ramzan et al. (2024) \cite{Ramzan2024} & Recommend crops using soil and environment data & IoT sensors + ML (KNN, SVM, RF, LightGBM, AdaBoost) & Local (Pakistan) & Combines real-time sensor data with static models \\
\hline

Soil Fertility & Abdellatif et al. (2021) \cite{Abdellatif2021} & Create soil fertility maps & GIS + Sentinel-2 + DEM + Kriging (DSQM model) & Regional (Egypt) & Uses multiple indices: fertility, chemical, geomorphologic \\
& Peng et al. (2022) \cite{peng_new_2022} & Predict fertility using crop spectral data & XGBoost + Backpropagation Neural Network & Field-scale (China) & Combines ML with vegetation indices (NDVI, VIF) \\
& Folorunso et al. (2023) \cite{folorunso_exploring_2023} & Review soil nutrient prediction models & Systematic review of ML models & Global & Highlights challenges in biological indicators and uncertainty \\
& El Behairy et al. (2024) \cite{el_behairy_accurate_2024} & Predict SQI & ANN (no GIS) using 16 soil attributes & Local (Drylands) & Pure ML with lab-based data, no spatial mapping \\
& Ben Abderrahmane et al. (2025) \cite{mine_comparative_2025} & Compare ML regressors for SQI & ML + SoilGrids data (4 models) & European & Produces soil quality maps using regression outputs \\
\hline

Soil Contamination & Taşpınar et al. (2021) \cite{Turky_soil_2021} & Assess heavy metal pollution & Field sampling + PCA + Enrichment Factor + Igeo & Local (Turkey) & Multivariate analysis of metal contamination \\
& Hammam et al. (2022) \cite{Egypt_contamnation_2022} & Map metal contamination spatially & GIS + Kriging + PCA + CF & Regional (Egypt) & Combines interpolation with contamination indices \\
& Diana et al. (2022) \cite{Malysia_Contamination_soil_2024} & Evaluate heavy metal contamination in soils & Contamination indices reading & Local (Malaysia) & No predictive or mapping applied \\
& Aparisi-Navarro et al. (2025) \cite{Portugal_contaminationAgroSoil_2025} & Model pollution risk spatially & GIS + Probabilistic Pollution Index (8 parameters) & Regional (Portugal) & Introduces PPI for pixel-wise risk classification \\
\hline

Soil Salinization & He et al. (2023) \cite{he_spatiotemporal_2023} & Predict future salinization trends & Remote sensing + RF, LightGBM, XGBoost & Regional (China) & Projects soil salinity up to 2050 \\
& Li et al. (2024) \cite{li_optimized_2024} & Monitor salinity changes over time & SVM and CART on Landsat imagery & Oasis-scale (China) & Assesses classification accuracy for salinity maps \\
& Hagage et al. (2024) \cite{hagage_monitoring_2024} & Soil Salinization and Waterlogging Monitoring & Remote sensing + IDW & Local (Egypt) & Assesses classification accuracy for salinity maps \\
\hline

Land Use Land Cover & Rohitha et al. (2023) \cite{shakya_land_2023} & Track land use change & PCA + K-means + MAD algorithm on satellite data & Local (Singapore) & Monitors water scarcity and land dynamics \\
& Gubila et al. (2024) \cite{gubila_land_2024} & examine LULC change & Landsat + ANOVA + MLC & Regional (Ethiopia) & mapping categorical and temporal LULC from 1986 to 2018  \\
& Brito et al. (2025) \cite{brito_land_2025} & examine LULC change & MapBiomas RS + Landsat maps + descriptive statistics & Regional (Brazil) & spacio-temporal mapping   LULC changes during 1985, 1990, 1995, 2000, 2005, 2010, 2015, 2020, 2022  \\
& K\"uç\"uk and Aslan (2025) \cite{kucuk_analysis_2025} & spatio-temporal LULC changes & CORINE land-cover data + GIS + descriptive statistics & Regional (Turkiye) &  describe LULC changes for the years 1990, 2000, 2006, 2012, and 2018.\\
& Senadi et al. (2025) \cite{senadi_spatio-temporal_2025} & Map urban/agricultural land changes & Landsat + Sentinel 2 + Google Earth Engine + RF + CART & Local (Algeria) & 40-year LULC analysis in urbanization context \\
\hline

Land Degradation & Oroud and Alghababsheh (2023) \cite{Oroud2023} & Land degradation changes & Landsat + NDVI + albedo  & Local (Jordan) & used statistical trend analysis (Mann–Kendall and Sen’s slope) and GIS-based spatial analysis to evaluate land degradation trends from NDVI and albedo time series from 1998 to 2021 \\
& Blanche et al. (2025) \cite{Blanche2024} & Map degradation risk areas & Sentinel + NDVI + NDWI + GIS + MLC & Local (Cameroon) & land cover degradation and geomorphological impacts of artisanal and semi-mechanized mining using classical statistical classification\\
& Ali et al. (2025) \cite{Ali2025} & Map degradation risk areas & Multi-index model + GIS + Kriging & Regional (Egypt) & Combines topographic, vegetation, erosion indices \\
\hline

Climate Change & Gabriele et al. (2023) \cite{gabriele_combined_2023} & monitoring climate change related to land degradation to support landscape preservation and planning & RS + GIS + MEDALUS + NDVI & Regional (Italy) & Use of composite degradation indices (MEDALUS), NDVI change detection, scenario mapping, decision support integration and spatial gradients of degradation sensitivity\\
& Lisetskii et al. (2024) \cite{lisetskii_geoinformation_2024} & Analyze climate impact on soil formation & GIS + 60-year climate and soil data & Regional (Crimea) & Models soil typology changes over decades \\
& Yılmaz (2024) \cite{yilmaz_trend_2024} & To detect long-term trends in precipitation and temperature parameters & RS + GIS + statistical tests (MK, SS)  & Regional (Turkey) & spatial integration of climatic and land cover data to highlight climatic shifts and their spatial patterns \\
& Piyasena and Bandara (2024) \cite{piyasena_utilizing_2024} & map and assess climate change vulnerability by combining exposure, sensitivity, and adaptive capacity dimensions & CVI + GIS + Multi-Criteria Decision Analysis & Local (Sri Lanka) & Multi-criteria geospatial vulnerability mapping, identification of highly vulnerable subregions and integration of socio-economic and biophysical data layers \\
& Lafortezza et al. (2024) \cite{lafortezza_geores_2024} & develop GEORES project to aid resilience and sustainability to climate change in urban zones & remote sensing + GIS + Statistical Analysis + ML & Regional (Italy) & monitoring of urban environmental parameters, supporting resilience planning; bridging research and application in urban climate resilience \\
\hline

\end{longtable}
\end{landscape}

\textcolor{white}{123}\\

\noindent Various case studies and several consistent themes have been identified. Applications related to geospatial soil quality consistently use machine learning techniques, remote sensing technologies, and Geographic Information Systems (GIS) to effectively address critical challenges in agriculture, land conservation, and environmental monitoring. Many of these studies utilize integrated approaches that combine in-situ data with satellite imagery and predictive modeling to improve accuracy and spatial resolution. However, there are still notable challenges and limitations within this field. The following section will outline existing obstacles and barriers that need to be addressed to improve system accuracy, robustness, and broader adoption.




\section{Challenges and Limitations}\label{challenges}
Despite advancements in soil quality analysis, several challenges and limitations remain. Even though geospatial techniques, machine learning, remote sensing, and digital soil mapping have greatly advanced, they still face substantial challenges. The following highlight some of the key issues documented in recent literature.

Many regions, especially in developing countries, suffer from inadequate field \textit{data}, which poses significant challenges for training and validating models \cite{folorunso_exploring_2023, DeCaires2025}. \textit{Ground-truth data} in these areas are often sparse, outdated, or unevenly distributed, limiting the calibration and validation of models \cite{AppChallgsDSMAfrica2024}. Additionally, \textit{biological soil properties} such as microbial biomass, enzyme activity, and soil fauna are often underrepresented in DSM frameworks due to the complexity and cost of measurement \cite{DevSQAframework2025}. The problem is intensified by variations in the soil classification systems, measurement protocols, units and scales used in geographic regions and research studies, which complicate the \textit{integration and comparability of data} \cite{Lagacherie2025}. There is also a semantic gap in how terms like "soil quality" and "soil health" are conceptualized across disciplines \cite{DevSQAframework2025}. Furthermore, many models do not report or propagate prediction \textit{uncertainty}, making users unaware of the confidence levels in spatial outputs \cite{DeCaires2025}. The management, storage, processing, and visualization of large, multi-temporal, multi-source geospatial datasets can be computationally intensive, especially in \textit{data-poor or low-resource environments}, where challenges such as reproducible workflows, automation, and insufficient data infrastructure further complicate the process \cite{DeCaires2025, Liakos2022}. Machine learning models, which are frequently perceived as "black boxes," also face challenges related to \textit{limited interpretability}, impacting decision-makers' capacity to comprehensively understand the results \cite{DeCaires2025}. In addition, soil quality is dynamic, with changes driven by land management, erosion, and climate variability, but many soil quality maps are static snapshots. \textit{Regular updates} to these maps can be costly or restricted by a lack of up-to-date covariates and field data \cite{AppChallgsDSMAfrica2024}. Finally, soil data and tools are often not translated into user-friendly formats for local farmers or policymakers, reducing their accessibility and practical utility.    


Despite existing obstacles, continuous progress in data science, artificial intelligence, and participatory sensing methodologies provides new avenues to address current limitations. The subsequent section delineates emerging trends and proposes future research and development trajectories.

\section{Trends and Future Directions}\label{trends}
As geospatial soil quality analysis systems continue to develop, several promising opportunities present themselves to enhance both methodologies and applications. The following areas are particularly important for further research and development.

\textit{Dynamic spatio-temporal modeling} is critical for developing models that can account for time-series changes in soil conditions driven by factors such as climate events, land use changes, or management practices \cite{DeCaires2025}. Current research tends to focus heavily on physical and chemical indicators. Future research should focus on the integration of biological indicators \textit{biological indicators} such as microbial biomass, soil enzyme activities, and microbiome composition, as these can serve as early warning signals of soil health decline and help improve resilience and adaptive capacity \cite{DevSQAframework2025,folorunso_exploring_2023,DeFord2024}. 

Furthermore, the development of XAI techniques can enhance model transparency, thereby making decision-support systems more interpretable and accessible to stakeholders \cite{Minasny2024,DeCaires2025}. Participatory sensing and citizen science initiatives—such as mobile applications, community-based monitoring, and the use of low-cost sensors—provide valuable opportunities to expand data collection efforts. Additionally, cloud computing and big data integration are crucial for processing large-scale datasets in near real time, particularly in remote or resource-limited regions. At the policy level, designing systems that directly \textit{support sustainable land management} and agricultural extension programs is essential to enable more informed and effective decision-making \cite{DeCaires2025}. 

Furthermore, there is an increasing focus on the development of advanced methodologies to \textit{quantify uncertainty} in spatial models. Bayesian design criteria and copula-based techniques have demonstrated efficacy in reducing prediction errors. Future research should focus on optimizing these methodologies and investigating novel approaches to improve the precision and robustness of spatial predictions in soil science \cite{Portugal_contaminationAgroSoil_2025,DeCaires2025}.    



As the domain advances, these technological innovations are anticipated to transform methodologies for evaluating and managing soil quality. The concluding segment synthesizes primary findings and underscores the significance of geospatial soil quality assessment systems in promoting sustainable land management practices.

\section{Conclusion}\label{concl}
This paper contributes a proposed pipeline that consolidates geospatial, machine learning, and domain-specific knowledge into a unified system. It outlines the core components, available datasets, analytical tools, and methodological approaches, along with their practical applications. The modular pipeline proposed in this review serves not only as a synthesis of current capabilities but also as a blueprint for future soil information systems that are dynamic, transparent, and scalable. The integration of geographic information systems (GIS), remote sensing modalities, and machine learning techniques has the potential to substantially improve the scope, efficiency, and precision of soil quality evaluations and assessments.

The discussion of \textit{challenges and limitations} has highlighted ongoing barriers, including data gaps, biases in machine learning predictions, limited interpretability, insufficient uncertainty quantification, computational constraints, and usability issues for non-specialists. Overcoming these obstacles necessitates not only technical advancements but also a focus on standardization, reproducibility, and effective stakeholder engagement.

Future development efforts may focus on dynamic spatio-temporal modeling, the integration of biological indicators, progress in XAI and uncertainty quantification, and the integration of Internet of Things (IoT) sensor networks with cloud-based GIS platforms. Furthermore, adopting participatory approaches and developing policy-oriented tools are essential to ensure geospatial soil quality systems are user-centric and directly relevant to sustainable land management practices.

These findings indicate that future geospatial soil quality assessment frameworks are expected to become increasingly dynamic, biologically integrated, interpretable, and tailored to the specific needs of farmers, policymakers, and researchers. Advancing along this trajectory will facilitate the transition of soil quality mapping from static diagnostic methodologies to proactive, data-driven decision support systems that improve agricultural resilience and advance environmental sustainability.


\subsection*{Funding Declaration} The authors declare that they received no funding for this work.

\bibliographystyle{ieeetr}
\bibliography{Refs,soilContamination,LandUseLCoverChange,SoilFertilityMapping,PredictionAgriculture,SoilSalinization,LandDegradation,ClimateChanges,Challenges}

\end{document}